\documentclass[jap,graphicx, reprint]{revtex4-1} % for checking your page length
\usepackage{amsfonts, amsmath, graphicx}
\usepackage[version=3]{mhchem}

\begin{document} 
\title{Curvature induced out of plane spin accumulation} 

\author{Zhuo Bin Siu}
\affiliation{Computational Nanoelectronics and Nanodevices Laboratory, Electrical and Computer Engineering Department, National University of Singapore, Singapore} 
\author{Mansoor B. A. Jalil} 
\affiliation{Computational Nanoelectronics and Nanodevices Laboratory, Electrical and Computer Engineering Department, National University of Singapore, Singapore} 
\author{Seng Ghee Tan} 
\affiliation{Data Storage Institute, Agency for Science, Technology and Research (A*STAR), Singapore} 

\begin{abstract}
In this work we show that (real space) curvature in the geometry of curved waveguides with Rashba spin orbit interaction (RSOI) can lead to out of plane spin accumulations. We first derive the RSOI Hamiltonian on arbitrarily curved surfaces. We then analyze the effects of curvature with two distinct methods. We first apply an adiabatic approximation on gently curved, planar waveguides lying flat on the $xy$ plane to show that analogous to the acceleration of the charge carriers by an electric field, the change in momentum direction of the charge carriers as they move along the waveguide leads to an out of plane spin accumulation. We then use the Heisenberg equations of motion to establish the relationships between spin currents and accumulations on non-planar waveguides. These relations predict the existence of out of plane spin accumulation on asymmetrically curved, non-planar waveguides. We finally solve for the eigenstates on such waveguides numerically, and present numerical results to verify our earlier analytic predictions. 
\end{abstract} 

\maketitle

\section{Introduction} 

The coupling between the spin and momentum degrees of freedom in systems with spin orbit interactions (SOI) allow the control of one to manipulate the other. In particular, it offers the intriguing possibility of manipulating spin by making charge carriers move through curved waveguides. The effective SOI fields, and hence spin accumulation will then vary with the momentum direction of the charge carriers as the latter move along the curved trajectories. 

Geometrically curved two-dimensional electron gas systems (2DEGs) with SOI have been studied by various authors.  	The earliest workers to study curved systems with SOI are Entin and Magarill who, as early as 1996 \cite{JETPLett64_460}, studied the effects of magnetic fields on cylinders with Rashba SOI (RSOI) \cite{JETP86_771, JETP88_815}. Trushin and Chudnovskiy proposed in 2006 the utilization a 1D curved RSOI arc as a spin switch \cite{JETPlett83_318}. This was followed shortly after by Zhang \textit{et al} who obtained analytic expression for the transmission through 1D RSOI curves of a few simple shapes \cite{PRB75_085308}. Transport through a RSOI cylinder without magnetic fields or magnetization sandwiched between two non magnetic leads were also studied in 2010 \cite{PRB81_075439, JAP108_033715}. The eigenstates of a RSOI cylinder with a constant magnetic field perpendicular to the cylinder axis were later obtained numerically \cite{JPhyConf193_012019}. Spin precession on RSOI cylinders \cite{PRB83_115305} and arbitrary one-dimensional paths on curved geometries \cite{PRB74_195314, PRB84_085307} have also been studied. We have previously proposed the use of a finite-width semi-circular arc with RSOI to generate out of plane spin polarized currents \cite{JAP115_17C513}. 

In this paper, we show analytically that the curvature of the waveguides can lead to out of plane spin accumulations absent in straight or flat SOI systems. We first present a derivation of the Hamiltonian for the RSOI on a arbitrarily curved surface where the coupling between spin and momentum, and hence the curved trajectory become evident from differential geometry arguments. Using this Hamiltonian, we then analyse the effects of curvature on the spin accumulation via two distinct methods. First, we use an adiabatic approximation method first used by some of us to explain the out of plane spin accumulation in the Spin Hall effect in SOI systems in the presence of an electric field \cite{NJP12_013016,JPSJ82_094714, SciRep5_18409}.  We show analytically that curvature on a curved waveguide lying flat on the $xy$ plane (Fig. \ref{grsCurved} ) can have an analogous effect to the electric field in the SHE. We refer to such curved waveguides lying flat on the $xy$ plane as `planar' waveguides.   Next, we then show that the constraints imposed by the Heisenberg equations of motion for the eigenstates of a time-independent Hamiltonian to derive the relations between the spin current and spin accumulations. These relations predict the presence of an out of plane spin accumulation on non-planar waveguides (Fig. \ref{gZcurv1var} ) that do not lie flat on the $xy$ plane.  We finally present numerical results to verify our predictions.  

\section {RSOI Hamiltonian on a curved surface} 

We first present a new  derivation of the RSOI Hamiltonian on an arbitrarily curved 2D surface described by the coordinates $q^1$ and $q^2$.  Here, we adopt the standard tensor analysis notation of denoting covariant (contravariant) quantities with subscript (superscript) indices.  For instance, a point on the surface of a cylinder of fixed radius $R$ with its axis lying along the $z$ direction may be described by the standard cylindrical coordinates $\phi$ and $z$ as $\vec{r} = R( \cos(\phi)e_x + \sin(\phi)e_y) + z e_z$, and we may identify $q^1 \rightarrow \phi$ and $q^2 \rightarrow z$.  

Our derivation hinges upon the difference between the action of a covariant derivative on a vector and a scalar, which we illustrate with the $(p)^2 / (2m)$ kinetic energy term.

The kinetic energy operator acting on a wave function $\psi$ on a Riemannian manifold can be written as 
\begin{equation} 
	\frac{1}{2m} p_ip^i\psi. \label{T0}
\end{equation}  
Despite their similar appearances, the $p^i$ to the immediate left of $\psi$ and the $p_i$ to the left of $p^i$ in Eq. \ref{T0} have different forms. The momentum operator $p_i$ acting on a wave function to its right on a Riemannian manifold is defined  as $p_i\equiv-i\nabla_i$. Here, $\nabla$ is the covariant derivative defined via
\begin{eqnarray}
	\nabla_i\psi&=&\partial_i\psi, \label{T11}\\
	\nabla_i v^j&=&\partial_iv^j+\Gamma^j_{ik}v^k \label{T12} 
\end{eqnarray}  
for a scalar $\psi$ in Eq. \ref{T11} and the $j$th contravariant component of the vector $v$ in Eq. \ref{T12}.  The $p^i$ on the right of Eq. \ref{T0} acts on the \textit{scalar} $\psi$ and gives the $i$th component of the vector $-i g^{ij}\partial_j\psi$ ; the $p_i$ to its left then acts on this $i$th component of the \textit{vector} to give 
\begin{eqnarray*}
	p_ip^i\psi &=& - \nabla_i(g^{ij} \partial^j\psi) \\
	&=&- (\partial_i+\frac{1}{2}\partial_i\ln(g))(g^{ij}\partial_j\psi) \\
	&=&- \frac{1}{\sqrt{g}}\partial_i(g^{ij}\sqrt{g}\partial_j\psi).
\end{eqnarray*}
In going from the first to second line we made use of the identity $\Gamma^i_{ij}=\frac{1}{2}\partial_j\ln g$, and from the second to the third line we recognized that $(\partial_i g^{ij}\partial_j+g^{ij}\partial_i\partial_j)\psi = \partial_i (g^{ij}\partial_j\psi)$. The last line contains the form of the Laplacian operator commonly seen in differential geometry textbooks. 

Returning to the RSOI effect, a traditional way of deriving the RSOI Hamiltonian on a curved waveguide lying flat on the $xy$ plane is to perform coordinate transformation from the corresponding RSOI Hamiltonian in Cartesian coordinates 
\begin{equation} 
	H_{RSOI} = \alpha(\vec{p}\times\hat{z})\cdot\vec{\sigma}=\alpha (p^y\sigma_x - p^x\sigma_y) \label{HRCa} 
\end{equation}
into the curvilinear coordinates $(q^1, q^2)$. Using the standard rules of tensor coordinate transformations $v^a = \frac{\partial x^a}{\partial{x'^i}}v^i$ and $v_a=\frac{\partial x'^i}{\partial x_a}v_i$, Eq. \ref{HRCa} becomes
\begin{equation}
	\alpha (\sigma_xp^y-\sigma_yp^x) = \alpha (\sigma_x\partial_i y-\sigma_y\partial_i x)p^i. \label{HR2}
\end{equation} 

It is desirable to obtain an expression for the RSOI Hamiltonian in terms of the spin operators along the tangential directions of the $q^1$ and $q^2$ coordinates  $\sigma_1 \equiv \partial_1 x^a\sigma_a$, $x^a=(x,y)$, and $\sigma_2$ defined analogously,  instead of the spin $x$ and $y$ operators. This can be accomplished by appealing to the fact that the RSOI Hamiltonian on the $xy$ plane is always given by $\alpha (\vec{p}\times\vec{\sigma})\cdot\hat{n}$ regardless of the basis vectors used to represent $\vec{p}$ and $\vec{\sigma}$.

Let $e_1$ and $e_2$ be the in-plane tangent vectors to a surface whose normal direction $e_3$ is defined as $e_1\times e_2 = |n|\hat{e}_3$. $e_1$ and $e_2$ may, in general, be position dependent. For example, the basis vectors $\hat{\phi}$ and $\hat{r}$ in standard cylindrical coordinates are dependent on $\phi$. It can be readily seen from $|n|^2 = |e_1|^2 |e_2|^2 (1- (\frac{e_1\cdot e_2}{|e_1||e_2|})^2)$ that $|n|=\sqrt{g}$.  We then have $ e_1\times\hat{e}_3 = \frac{1}{\sqrt{g}} e_1\times(e_1\times e_2)  = \frac{1}{\sqrt{g}} e_1g_{12} - e_2 g_{11}  = -\sqrt{g} e^2$, and similarly, $e_2\times\hat{e}_3 = \sqrt{g} e^1$. This gives 
\begin{equation}
	\alpha\vec{p}\times\hat{n} =  \alpha ((\sqrt{g}\sigma^1)p^2 - (\sqrt{g}\sigma^2)p^1) \equiv S_1 p^1 + S_2 p^2 \label{SIdefn}
\end{equation}
where we defined the $S_i$s. Comparing Eq. \ref{SIdefn} with Eq. \ref{HR2} gives $\alpha (\sigma_x\partial_i y - \sigma_y\partial_i x) = S_i$. We note that we could also have written the above as $S^ip_i$ since $S_i p^i = S^i g_{ij} p^j = S_j p^j$. We use both forms interchangeably. 

Acting on a wave function to the right, we have
\begin{equation}
	S_ip^i\psi=-i (S_ig^{ij})\partial_j\psi. \label{RSOI2}  
\end{equation}

It is perhaps a bit surprising that despite the appearance of the $S_i$s only to the left of the $\partial_i$s where each of the $S_i$s may be position dependent, the expression $S_ip^i$ \textit{summed over the $i$ coordinates} is Hermitian since this is just a restatement of the manifestly Hermitian $\alpha(p^x\sigma_y-p^y\sigma_x)$. We explain this unexpected result later in this section when we move on to a more general derivation of the RSOI Hamiltonian valid even for non-planar curved surfaces that do not lie flat on the $xy$ plane.  It should be noted, however, that the $i$th component of $S_ip^i$ by itself, \textit{without} summing over $i$, is in general not Hermitian. 

We shall next exhibit a more symmetric form of the RSOI Hamiltonian which is applicable even when the surface normal vector $\hat{n}$ is no longer the constant $\hat{z}$ considered earlier, but varies in space for a non-planar surface.

We start off by writing the position-momentum symmeterized RSOI Hamiltonian as  
\begin{equation}
	\frac{1}{2}\{p_i,S^i\}. \label{RSOI1}
\end{equation}

Although the form of Eq. \ref{RSOI1} differs slightly from that Eq. \ref{RSOI2}, we shall show that the former reduces to the latter.   As a practical prescription to writing down the Hamiltonian on a curved surface explicitly,  Eq. \ref{HR2} differs from Eq. \ref{RSOI2}. In the former, the prescription is to first write $p^x$ and $p^y$ into the $q^i$ coordinates, and then to treat the resulting factors that multiply $p^1$ and $p^2$ as $S_1$ and $S_2$ respectively. For Eq. \ref{RSOI1},  we write down the $S^i$s directly based on their definition in Eq. \ref{SIdefn}. 

The key to our derivation of the RSOI Hamiltonian is to again note that the second $p_i$ in $\{ S^i, p_i\}\psi= (p_iS^i + S^ip_i)\psi$ nearer $\psi$ acts on the scalar $\psi$ and does not have the Christoffel symbols, while the first $p_i$ further away from $\psi$ acts on the vector with \textit{vector components} $S^i\psi$, and contains the Christoffel symbols. Expanding, we thus have 
\begin{eqnarray}
	\frac{1}{2}(p_iS^i + S^ip_i)\psi &=&-\frac{i}{2}  \big((\partial_i+\frac{1}{2}\partial_i\ln g)S^i + S^i\partial_i\big)\psi \nonumber \\ 
	&=&- \frac{i}{2}(\mathrm{div}\ \vec{S})\psi + S_ip^i \psi. \label{RSOI3}
\end{eqnarray}

Eq. \ref{RSOI3} contains Eq. \ref{RSOI2} plus a term proportional to $\mathrm{div}\ \vec{S}=\mathrm{div}\ (\hat{n}\times\vec{\sigma})$ which is really $\vec{\nabla}\times{\hat{n}}\cdot\vec{\sigma}$. This can be interpreted as the commutator arising in $\frac{1}{2}\{ S^i,p_i\} = S^ip_i + \frac{1}{2}[p_i, S_i]$.

We stress that the form of Eq. \ref{RSOI3} differs from that which comes from a naive expansion of $\frac{1}{2} (S^i p_i + p_i S^i)  \stackrel{!}{=} (S^i p_i + \frac{1}{2}[p_i,S^i]) \stackrel{!}{=} (S^ip_i + \frac{1}{2}i\partial_i S^i)$ which does not capture the $\frac{1}{2}(\partial_i\ln g) S^i$ terms in Eq. \ref{RSOI3}.  The $\frac{1}{2}(\partial_i \ln g)$ terms are crucial for Eq. \ref{RSOI3} to be Hermitian in the presence of curvature. 

There is a further simplification to Eq. \ref{RSOI3}. It turns out that $\mathrm{div}\ \vec{S}$ term is identically 0. After some algebra, $S^1$ and $S^2$ can be shown to take the rather simple forms  
\begin{eqnarray}
	S^1 &=& -\frac{\alpha}{\sqrt{g}}\sigma_2, \\
	S^2 &=& \frac{\alpha}{\sqrt{g}}\sigma_1.
\end{eqnarray}
 Using the above and the fact that $\partial_2 e_1 = \partial_1 e_2 = \partial_1\partial_2 \vec{r}$, we have 
\begin{eqnarray*}
	\partial_1 S^1 + \partial_2 S^2 &=& -\frac{1}{\sqrt{g}} ((-\partial_1 \ln\sqrt{g}) \sigma_2 + (\partial_2 \ln \sqrt{g}) \sigma_1)\\
	&=&-(\partial_1 \ln\sqrt{g}S^1 + \partial_2\ln\sqrt{g}S^2),
\end{eqnarray*}
so that $\nabla\cdot\vec{S} = (\partial_i + (\partial_i \ln\sqrt{g}))S^i = 0$ identically. This explains why the RSOI Hamiltonian can be written as $S^ip_i$ with the momentum operators to the right of the, in general, position dependent spin operators but yet remain Hermitian.     

We thus close this section by summarizing that the SOI Hamiltonian on a curved surface acting on a wavefunction is given by 
\[ 
	H_{RSOI}\psi  = -\frac{i\alpha}{\sqrt{g}} (-\sigma_2\partial_1 + \sigma_1\partial_2)\psi 
\]

\section{Out pf plane spin accumulation} 
Having derived the Hamiltonian for the RSOI on a curved surface, we now use this Hamiltonian to analyze the effects of curvature on the spin accumulation via two distinct techniques. 

In the first technique, we invoke the adiabatic approximation in which we assume that the geometric curvature is gentle enough that the spin of charge carriers adiabatically follow the direction of the local SOI field. We show that the change in momentum of the charge carriers moving along the curved trajectory has an analogous effect to the acceleration of the charge carriers by an electric field in the spin Hall effect and leads to an out of plane spin accumulation.  

In the second technique, we make use of the fact that since the expectation values of observable quantities for the eigenstates of a time independent Hamiltonian do not very in time, the Heisenberg equation of motion for these observables imposes constraints between the expectation values of the various observables that appear in these equations of motion. We show that in particular, these constraints predict the existence of finite out of plane spin accumulation integrated across the width of an asymmetrically curved, infinitely long waveguide. 

\subsection{Adiabatic approximation}\label{sectAdApprox}
Aharonov and Stern first noted in the early 90s \cite{PRL69_3593} that a magnetic particle whose magnetization adiabatically follows a strong, time varying in-plane magnetic field gains an out of plane magnetization. This idea was later exploited by Fujita and co-authors \cite{NJP12_013016, JPSJ82_094714, SciRep5_18409} to explain the out of plane spin accumulation in the spin Hall effect. There, the role of the time varying magnetic field is played by the RSOI field, and the time variation of the field caused by an applied electric field. The acceleration in the direction of the electric field causes the momentum direction to rotate towards the direction of the field. The direction of the RSOI field is perpendicular to that of the momentum, and rotates along with the momentum.  We show here that instead of applying an electric field, the change in the momentum direction of a particle as it moves along a curved path can give rise to an analogous effect.

\begin{figure}[ht!]
\centering
\includegraphics[scale=0.5]{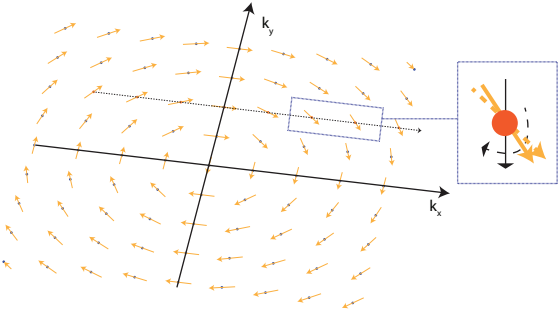}
\caption{ The SOI assigns an effective magnetic field to every point in $k$ space. The dotted line shows the trajectory of a charge carrier in $k$ space as it is accelerated in the $x$ direction by an electric field. The inset shows the rotation of the carrier spin, and the effective torque acting on the spin, as it moves between two neigbouring points in $k$ space.  } 
\label{gA2SOIfield}
\end{figure}

Fujita considered a flat 2DEG RSOI system subjected to an electric field, which without loss of generality we take to be in the $x$ direction. The $p^2/2m$ kinetic energy term is not central to the argument, and we shall ignore it. The relevant terms are then $\vec{B}(\vec{k})\cdot\vec{\sigma} + Ex$ where we wrote $\vec{B}(\vec{k}) = \alpha(k_y,-k_x,0) = k\hat{b}$ to stress that this takes the form of a momentum dependent effective magnetic field. Putting the $Ex$ electric field term aside, $\vec{B}(\vec{k})$ can be thought of as assigning of a magnetic field to every point in $k$ space, much like a magnetization $\vec{M}(\vec{r})\cdot\vec{\sigma}$ assigns a magnetization vector to every point in real space, as shown in Fig. \ref{gA2SOIfield}.  We next consider what happens when we now include the $x$ electric field. Between successive moments in time, the electric field increases $k_x$ while $k_y$ remains unchanged. The direction of the unit vector $\hat{k}$ changes, and we may think of the charge carrier as adiabatically tracing out a path in the $k_x$ direction in $k$ space as shown in the figure. The direction $\vec{B}(\vec{k})$ changes at different points on the trajectory. In the adiabatic approximation we assume that the spin of the charge carriers follows that of the changing $\vec{B}(\vec{k})$ so the spin rotates along with $\vec{B}(\vec{k})$. This rotation can be interpreted as being due to a spin torque in the out-of-plane direction, whose mathematical form we shall now proceed to exhibit.

We define a unitary transformation $U$ which satisfies 
\begin{eqnarray*}
	U\hat{b}\cdot\vec{\sigma}U^\dagger &=& \sigma_x ,\\
	U(\partial_{k_x}\hat{b}\cdot\vec{\sigma}) U^\dagger &=&  |\partial_{k_x}\hat{b}|\sigma_y. 
\end{eqnarray*}

 With these properties, we can deduce that 
 \begin{eqnarray*}
 	&& \partial_{k_x} (U\hat{b}\cdot\vec{\sigma}U^\dagger) = \partial_{k_x}\sigma_x = 0 \\
	&\Rightarrow& (\partial_{k_x} U)\hat{b}\cdot\vec{\sigma}U^\dagger + |\partial_{k_x}\hat{b}|\sigma_y + U\hat{b}\cdot\vec{\sigma}(\partial_{k_x} U^\dagger) = 0 \\
	&\Rightarrow& (\partial_{k_x}U)U^\dagger\sigma_x+ |\partial_{k_x}\hat{b}|\sigma_y + \sigma_x U(\partial_{k_x}U^\dagger) = 0 \\
	&\Rightarrow&[(\partial_{k_x}U)U^\dagger,\sigma_x] = -|\partial_{k_x}\hat{b}|\sigma_y.
 \end{eqnarray*}
In going from the first to second line we made use of $U(\partial_{k_x}\hat{b}\cdot\vec{\sigma}) U^\dagger = \sigma_y$;  from the second to third line we inserted a $1=U^\dagger U  = U U^\dagger$ into the appropriate places and then used $U\hat{b}\cdot\vec{\sigma}U^\dagger = \sigma_x$, and from the third to the fourth line the identity $(\partial_x U^\dagger)U = - U^\dagger\partial_x U$.  From the last line, we can deduce that $(\partial_{k_x}U)U^\dagger = i|\partial_{k_x}\hat{b}|\sigma_z$.

We now perform the unitary transformation on the RSOI Hamiltonian and the electric field
\begin{eqnarray*}
	U (\vec{B}\cdot\sigma + Ex) U^\dagger &=& |B|\sigma_x + E (x + [U,x]U^\dagger) \\
	&=&|B|\sigma_x   + E(x - i(\partial_{k_x}U)U^\dagger) \\
	&=&|B|\sigma_x + E(x + |\partial_{k_x}\hat{b}|\sigma_z)
\end{eqnarray*}

Since $\hat{z}=\hat{x}\times\hat{y}$ we identify the $\sigma_z$ term in the transformed frame as pointing in the $\hat{b}\times(\partial_{k_x}\hat{b}) = \pm \hat{z}$ in the the original frame where the $+$ or $-$ sign depends on the specific directional relationship between $\hat{b}$ and $\partial_{k_x}\hat{b}$. This $\sigma_z$ term here can be interpreted to serve two roles. First, it results in an out-of-plane spin accumulation for the carriers. Second, it provides the torque required for the spin to rotate in the in-plane direction and adiabatically follow the direction of $\vec{B}\cdot\vec{\sigma}$, as can be readily verified by using Heisenberg equation of motion on $i\partial_t |\vec{k}\rangle\sigma_z\langle \vec{k}|$ for a particle polarised in the original frame $\hat{b}$ / transformed frame $\hat{x}$ direction. 

\begin{figure}[ht!]
\centering
\includegraphics[scale=0.7]{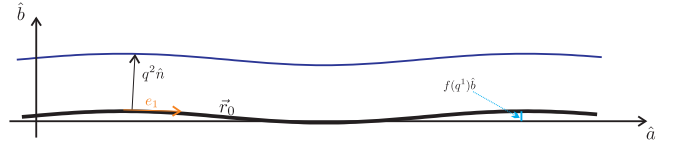}
\caption{$\hat{a}$ and $\hat{b}$ are orthogonal vectors lying on the $xy$ plane with $\hat{z}=\hat{a}\times\hat{b}$. The thick black line represents the curve traced out by $\vec{r}_0$ and the thin blue line the locus of points with constant $q^2$. We consider $\vec{r}_0 = q^1 \hat{a} + f(q^1)\hat{b}$ where $f(q^1)$ is small so that the waveguide is largely oriented along the $\hat{a}$ direction.   } 
\label{gIPgraph}
\end{figure}

A similar analogy can be made for a gently curved 2DEG lying flat on the $xy$ plane. Consider such a curve parameterized by 
\begin{equation} 
	\vec{r} = \vec{r}_0(q^1) + q^2 \hat{n}(q^1)
	\label{eqCwg1}
\end{equation}

 where $\hat{n} \propto (\partial_1{\vec{r}_0}\times\hat{z})\Big|_{q^2=0}$. Fig. \ref{gIPgraph} shows one example of a curve with a small curvature oriented largely along an arbitrary direction denoted as $\vec{a}$. We assume that the curvature is small enough that momenta of the  carriers adiabatically follow the curvature, and that the spin of the particles constrained to move along such a curve, for example due to the curvature of the waveguide carrying the particles, adiabatically follows the position dependent direction of the SOC field as well.

We then have
\begin{eqnarray*}
	e_1 &=& \partial_1 \vec{r}_0(q^1) + (\partial_1 \hat{n})q^2 \\
	e_2 &=& \hat{n}.
\end{eqnarray*}
Note that the requirement that $\partial_1 \hat{n}$ is perpendicular to $\hat{n}$ and that it is perpendicular to $\hat{z}$ lead to $\partial_1 \hat{n}$ being parallel (or antiparallel) to the tangent vector $\partial_1\vec{r}_0$ at $q^2=0$. This leads, in turn, to $g=g_{11}$ and 
\[ 
	g_{ij} = \begin{pmatrix} g_{11} & 0 \\ 0 & 1 \end{pmatrix}, g^{ij} = \begin{pmatrix} \frac{1}{g} & 0 \\ 0 & 1 \end{pmatrix}.
\]

We now introduce the unitary transformation $U$ so that $U\sigma_2U^\dagger = \sigma_y$, $U(\partial_1 \sigma_2)U^\dagger = -|\partial_1 e_2|\sigma_x$. Proceeding in a similar fashion as before, we have $(\partial_1 U)U^\dagger = i|\partial_1 e_2|\sigma_z$ where again, $\sigma_z$ in the transformed frame points along the original frame $\pm z$ direction.

Noting that $e_1\propto \partial_1\hat{n}$, we may write $U\sigma_1U^\dagger = c_1\sigma_x$ where $c_1 = e_1\cdot\partial_1\hat{n}$. Applying the unitary transformation $U$ on the RSOI Hamiltonian and recalling that $S^1=\frac{\alpha}{\sqrt{g}}(-\sigma_2)$,$S^2=\frac{\alpha}{\sqrt{g}}\sigma_1$,

\begin{eqnarray*}
	&& U(\{p_i,S^i\}U^\dagger / \alpha \\
	 &=&  (p_1-|\partial_1 e_2|\sigma_z)(-\frac{\sigma_y}{\sqrt{g}}) + (-\frac{\sigma_y}{\sqrt{g}})(p_1-|\partial_1 e_2|\sigma_z) \\
	 &\ & + \{ p_2, \frac{c_1 \sigma_x}{\sqrt{g}}\} \\
	&=&(-\{p_1,\frac{\sigma_y}{\sqrt{g}}\}+\{p_2,\frac{c_1 \sigma_x}{\sqrt{g}}\}).
\end{eqnarray*}
In going from the first to second lines we made use of the fact that $\{ \sigma_x,\sigma_y \}=0$. This shows that the products of the transformed spin operators, and the commutators between their respective momentum operators and the unitary transformation, cancel out. 

Performing the unitary transformation on the kinetic energy operator gives 
\[ 
	\frac{1}{2m} Up^2U^\dagger = \frac{1}{2m} ( (p_1-|\partial_1 e_2|\sigma_z)(p^1-g^{11}|\partial_1 e_2|\sigma_z) + p^2p_2 ) 
\] 
so that putting both the transformed RSOI and kinetic energy operators together gives 
\begin{eqnarray*}
	&& UHU^\dagger \\
	&=& \frac{1}{2m}\Big( (p_1-|\partial_1 e_2|\sigma_z-g_{11}\frac{m\alpha}{\sqrt{g}}\sigma_y)(p^1-g^{11}|\partial_1 e_2|\sigma_z\\
	&-& \frac{m\alpha}{\sqrt{g}}\sigma_y)\\
	&+& (p^2+\frac{mc_1\alpha}{\sqrt{g}}\sigma_x)(p_2+\frac{mc_1\alpha}{\sqrt{g}}\sigma_x) - \frac{(m\alpha)^2}{4g}(g_{11}+c_1^2) \Big).
\end{eqnarray*}
The curvature can thus be interpreted as introducing a gauge $A_i = (0, -|\partial_1\hat{n}|\sigma_z,0)$ that points in the spin out of plane $z$ direction. The presence of this term, similar to its analog in Fujita's works, leads to a finite spin polarization in the lab $z$ direction as well as provides the spin torque necessary for a spin adiabatically following the direction of the local SOC field to rotate along with the field as the field changes direction with $q^1$.

We can derive an approximate analytical expression for the spin $z$ polarization in a gently curved segment. Consider
\[
	\vec{r}_0 = q^1 \hat{a} + f(q^1) \hat{b}
\]
where $\hat{a}$ and $\hat{b}$ are orthogonal unit vectors representing the `average' longitudinal and transverse directions of current flow in the sense that $\partial_1 f$ can be taken to be small so that keeping only the linear terms in $\partial_1 f$ in the following is an adequate approximation. The definition of $U$ has been chosen so that $\sigma_z$ in the rotated frame points along the lab frame spin $+z$ direction for a positive $\partial_1^2 f$. 

We also assume the curve to be gently curved so that it is adequate to keep only the linear terms in $\partial_1^2 f$ and discard higher derivatives of $f$. $\vec{r}_0$ hence describes an approximately straight line along the $\hat{a}$ direction. 

Under these approximations,
\begin{eqnarray*}
	&&UHU^\dagger\\
	 &\approx&\frac{1}{2m} (1 + q^2(\partial_1^2f))((p_1)^2 - \{ |\partial_1^2f|, p_1\}\sigma_z + \\
	 &&  \frac{1}{4}|\partial_1\sigma_z|^2 + (p_2)^2) \\
	&+& \alpha(1-\frac{1}{2}q^2(\partial_1^2f))(\mathrm{sgn}(\partial_1^2f) p_2\sigma_x-p_1\sigma_y).
\end{eqnarray*}

We assume that $|\partial_1^2 f|$ is small enough that $p_1$ and $p_2$ are still approximately good quantum numbers so that the weightages of the spin polarizations along the spin $x$, $y$ and $z$ directions are proportional to the momenta multiplying the respective spin operators.  The out of plane spin $z$ polarization in the original frame is then given by 
\begin{equation} 
	\langle \sigma_z \rangle \approx \frac{ p_1 \partial_1^2 f}{m\alpha\sqrt{p_1^2+p_2^2}}.
	\label{approxSz} 
\end{equation}

\subsection{Heisenberg equation of motion} 

The results of the previous section are based on the adiabatic approximation. In this section, we use another approach to show the presence of out-of-plane spin accumulation due to curvature without invoking the approximation. 

The starting point of this approach is the observation that the time derivative of any observable for an energy eigenstate of a time independent Hamiltonian is zero. The absence of time dependence is, as we shall see later, the net result of quantities related to spin and charge currents and forces balancing off each other. While the concept of a spin force \cite{PRL95_187203} has been used to explain \cite{EPL107_37005,JPSJ82_094714} various spin related phenomena such as the spin Hall effect \cite{PLA35_459, PRL83_1834}, exploiting the fact that the expectation values of (divergences of) currents and forces cancel out for energy eigenstates is, to the best of our knowledge, a novel method of extracting the relationships between them. The primary tool for evaluating the local spin and charge currents will be the Heisenberg equation of motion (EOM). We show in the appendix that for an arbitrary observable $O$, 
\begin{eqnarray}
	-i[|O|\vec{r}\rangle  \langle \vec{r}|, p^2] &=& -\Big( \big((\partial_i + (\partial_i \ln\sqrt{g}))\{O|\vec{r}\rangle  \langle\vec{r}|, p^i \}\big) \nonumber \\
	&-& \{(\partial_i O) |\vec{r}\rangle\langle\vec{r}|,p^i\} \Big).
\label{KEeom}
\end{eqnarray}

We also have
\begin{eqnarray}
	&& -i \langle\psi| [O|\vec{r}\rangle\langle\vec{r}|, H_\mathrm{RSOI}] |\psi\rangle \nonumber \\
	&=& -i \Big( \psi^\dagger O(-iS^i\partial_i \psi) - (- iS^i\partial_i \psi)^\dagger O\psi \big) \nonumber  \\
	&=&-\frac{1}{2} \langle\psi|  i[|\vec{r}\rangle\langle\vec{r}|\{O, S^i\}, p_i] - \{ |\vec{r}\rangle\langle\vec{r}|(-i [O, S^i] ), p_i \} |\psi\rangle \label{symSOcomm1}.
\end{eqnarray}

In going from the second to the third line in Eq. \ref{symSOcomm1} we expanded $OS^i = \frac{1}{2} ( \{ O, S^i\} + [O, S^i])$ and analogously for $S^iO$. 

There is a potential point of confusion in going from the first to second line where one might naively believe that there should be terms proportional to $\partial_i S^i$ due to the $S_i$ and $p_i$ not commuting with each other. However, our arguments are based on the fact that the expectation values of eigenstates do not change in time, a fact that is rooted in
\begin{eqnarray*}
	&&\psi^\dagger O(\vec{r}) H(-i\nabla) \psi -  \psi^\dagger H(-i\nabla)^\dagger O(\vec{r})  \psi \\ 
	&=& \psi^\dagger O(\vec{r}) (H(-i\nabla) \psi) -  (H(-i\nabla)\psi)^\dagger O(\vec{r})  \psi \\
	&=& \psi^\dagger O(\vec{r}) \psi E - E \psi^\dagger O(\vec{r}) \psi \\
	&=& 0.
\end{eqnarray*}
Here, we wrote $O(\vec{r})$ and $-i\nabla$ to denote that the equations are in the position basis; the use of Hermitian conjugate $\psi^\dagger$ rather than simple complex conjugation $\psi^*$ in the above do not indicate second quantization but is due to the fact that the operators are two by two matrices and states two by one vectors in spin space.  We emphasize that the $-i\nabla$ momentum opertors in $H$ do not act on the operator $O$ anywhere to result in terms proportional to $\partial_i O$.

\subsubsection{Local spin and charge currents} 

We first derive the local charge current. Putting $O=\mathbb{I}_\sigma$ into Eq. \ref{symSOcomm1} gives
\begin{eqnarray*}
	-i[|\vec{r}\rangle\langle\vec{r}|, \frac{1}{2}\{S^i,p_i\} ] &=& =-i [S^i|\vec{r}\rangle\langle\vec{r}|, p_i] \\
	&=& -\partial_i (|\vec{r}\rangle\langle\vec{r}|S^i) + |\vec{r}\rangle\langle\vec{r}|(\partial_i S^i) \\
	&=& -(\partial_i + (\partial_i \ln\sqrt{g}) )(|\vec{r}\rangle\langle\vec{r}| S^i).
\end{eqnarray*}
where in going from the second to the third line we made use of the fact that $(\partial_i+(\partial_i \ln\sqrt{g}))S^i=0 \Rightarrow -\partial_i S^i = \partial_i \ln\sqrt{g}$  proven earlier and identify 
\[
 |\vec{r}\rangle \langle\vec{r}|S^i
\] as a contributory term to the charge current in the $e_i$ direction. There is an additional contribution to the charge current due to the kinetic energy $\frac{p^2}{2m}$ term. Applying Eq. \ref{KEeom}, the total charge current in the 2DEG reads 
\[ 
	j^i = \frac{1}{2} \{|\vec{r}\rangle\langle\vec{r}|,\frac{p^i}{m} \} + |\vec{r}\rangle S^i\langle\vec{r}|.
\] 

This is the `local' version (in the sense that it gives the current at specific position $\vec{r}$) of the current operator one would expect from the `global' expression $\vec{v} = -i[\vec{r}, H]$. 

We next consider the temporal evolution of the local spin accumulation.

Using Eq. \ref{symSOcomm1}, we have
\begin{eqnarray} 
	&&-i[|\vec{r}\rangle \sigma_i \langle\vec{r}|, \frac{1}{2}\{S^j,p_j\}]  \nonumber \\
	&=& -\frac{1}{2} \langle\psi|  i[|\vec{r}\rangle\langle\vec{r}|\{\sigma_i, S^j\}, p_j] \nonumber \\
	&& - \{ |\vec{r}\rangle\langle\vec{r}|(-i [\sigma_i, S^j] ), p_j \} |\psi\rangle.
\label{commSiSOC} 
\end{eqnarray}

A direct consequence of the definition of an eigenstate is that the time derivative of the expectation value of any local observable for the eigenstate of a time independent Hamiltonian must be 0. The Heisenberg equations of motion for the time derivatives of such expectation values can thus be interpreted as balance equations where the sum of all contributing terms must add up to 0. We have taken care during our derivation of Eqs.  \ref{KEeom} and \ref{symSOcomm1} to ensure that each of the individual terms on the right hand side are Hermitian so that each term corresponds to an observable quantity. The requirement that these expectation values sum up to 0 constraints how they are related to one another. 

Returning now to Eq. \ref{commSiSOC}, we note that the term on the right hand side has two parts. $\{ \sigma_i, S^j \}$ is proportional to the component of $S^j$ pointing in the $e_i$ direction. This and $[|\vec{r}\rangle\langle\vec{r}|, p_j]= i\hbar\partial_j (|\vec{r}\rangle\langle\vec{r}|)$ admit the interpretation of the  $\frac{1}{2} \{ \sigma_i, S^j\}[|\vec{r}\rangle\langle\vec{r}|, p_j]$ term without summing over $j$ as being proportional to derivative of the component of $S^j$ parallel to $e_i$ in the $j$ direction. The second term on the right hand side $[\sigma_i, S^j]\{ |\vec{r}\rangle\langle\vec{r}|, p_j \}$ can be interpreted as torque acting on $\sigma_i$ due to the local value of the SOI field $\frac{1}{2} \{ |\vec{r}\rangle  S^j \langle\vec{r}|, p^j \}$. (The anticommutator between two spin operators gives the dot product of the vectors in spin space the spin operators point to; their commutator gives $i$ times a spin operator pointing in the direction of their cross product. ) 

In a 2DEG the $p^2/2m$ kinetic energy term gives further contributions. Putting everything together,  we have 
\begin{eqnarray} 
	&&\partial_t |\vec{r}\rangle\sigma_i\langle\vec{r}| \nonumber \\
	&=& \frac{1}{2} \Big(  -(\partial_j+(\partial_j\ln g))  \{ \frac{1}{m}p^j, |\vec{r}\rangle\sigma_i\langle\vec{r}| \} + \{ |\vec{r}\rangle \partial_j \sigma_i\langle \vec{r}|, \frac{1}{m}p^j\}  \nonumber \\
	&& -i[|\vec{r}\rangle\langle\vec{r}|\{\sigma_i, S^j\}, p_j] + \{ |\vec{r}\rangle\langle\vec{r}|(-i [\sigma_i, S^j] ), p_j \} \Big). \label{hEOM}
\end{eqnarray}

The first term consists of the divergence of a quantity which admits the interpretation of a local spin $i$ current flowing in the $l$ direction $j_{\sigma_i}^l$, 
\[
j_{\sigma_i}^l \equiv \frac{1}{2}\{\frac{1}{m}p^l,|\vec{r}\rangle\sigma_i\langle\vec{r}| \}.
\]

The second term can be interpreted as a spin current correction term for the position dependence of the spin operator whose time derivative is being taken. The next two terms have been discussed previously. Taken together, the entire expression can be interpreted as saying that the spin torque acting on the local spin density due to the SOI effective magnetic field are balanced off by the divergence in spin current, and the spatial variation in the local spin density. We note in passing that the link between the divergence in spin current and the spin torque has also been pointed out earlier in Refs. \cite{SolStateComm139_31} and \cite{PRB77_075304}, in those case for spin torque resulting from an external magnetization. 

\subsubsection{Non planar curved SOI systems}

We now move on to study non-planar curved systems of the form $\vec{r} = (q^1, y(q^2), z(q^2))$. The $e_1$  direction is then simply the $x$ direction, while $e_2 = (0, \partial_2 y, \partial_2 z)$. To simplify the physical picture we restrict ourselves to $z$ having a dependence only on $q^1$ and not $q^2$.  For example, the $q^1$ coordinate can be the $x$ coordinate along the top half of a not necessarily circular cross section of a uniform, infinitely long cylinder, and $q^2$ the dimension along its length.  In such systems where one or more of the dimensions are curved, there is an additional energy contribution commonly known as the da Costa geometric confinement potential \cite{PRA23_1982, AnnPhy63_586} which can be physically interpreted as the energy contribution from the `force' confining the charged carriers to stay on the curved surface. This term has the mathematical form of  $-\kappa^2/(8 m)$ where $\kappa$ is the radius of curvature. While this potential does not affect the relations obtained from the Heisenberg equation of motion in this section,  it does affect the distribution of the charge density by favouring the concentration of charge into regions with smaller radaii of curvature.

Due to the translational invariance along the $q^2$ direction, the wavefunction of the eigenstate can be written in the form of $\Psi = \exp(i k_x q^1)\psi(q^2)$. 

We have 
\begin{eqnarray}
	&& \partial_t (|\vec{r}\rangle\langle\vec{r}|)  \nonumber \\
	&=& -\frac{1}{2} \Big( \partial_1    \{ \frac{p^1}{m} - \frac{\alpha}{\sqrt{g}}\sigma_2  , |\vec{r}\rangle\langle\vec{r}| \}  ) \nonumber \\ 
	&&+ (\partial_2+\partial_2 \ln \sqrt{g}) ( \{ \frac{p^2}{m} + \frac{\alpha}{\sqrt{g}}\sigma_1  , |\vec{r}\rangle\langle\vec{r}| \} ) \Big). \label{2degz1dtn} 
\end{eqnarray} 

\begin{figure}[ht!]
\centering
\includegraphics[scale=0.5]{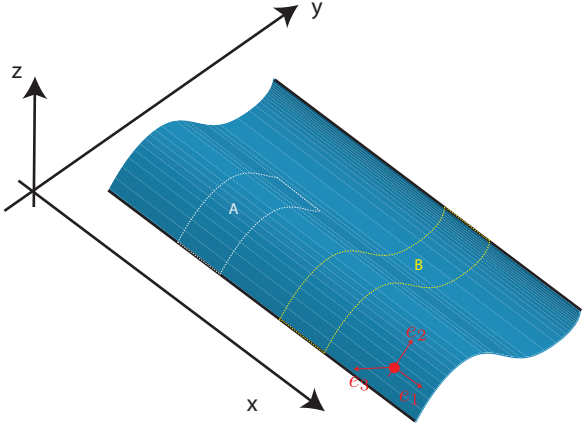}
\caption{The thick black lines denote the transverse edges of the waveguide which extends to infinity along the $x$ / $e_1$ direction. Both Gaussian pillboxes A and B have edges parallel to $e_1$ and $e_2$, and one edge parallel to $e_2$ extending slightly over the edge of the waveguide where the wavefunction and its derivatives go to 0. The remaining edge of pillbox B extends over the opposite edge, while that of pillbox A remains within the pillbox.  } 
\label{gZcurv1var}
\end{figure}

We consider $p_1$ eigenstates in the absence of a mangetization, and integrate the expectation value of both sides of Eq. \ref{2degz1dtn} over the Gaussian pillbox A in Fig. \ref{gZcurv1var}.  Translational invariance along the $e_1$ direction implies that the contributions to the integral from the two edges parallel to $e_1$ cancel out, while the edge outside the waveguide does not contribute. The fact that $\partial_t \rho = 0$ everywhere then implies that  
\[
	 \big(\langle p^2(\vec{r}) \rangle  + \frac{ m \alpha}{\sqrt{g}}\langle \sigma_1(\vec{r})  \rangle \big) = 0. 
\]
where $\langle p^2(\vec{r})\rangle \equiv \langle | (\frac{1}{2} \{ |\vec{r}\rangle\langle\vec{r}|, p^2 \} |\rangle$.  We can then conclude that 
\begin{equation}
	\langle \sigma_1(\vec{r}) \rangle = -\frac{1}{m\alpha} \langle p^2(\vec{r})\rangle \label{s1p2Eq}
\end{equation}  
as we might have expected from the form of the Hamiltonian.

We consider the time evolution of $|\vec{r}\rangle\langle\vec{r}|\sigma_1$. Using Eq. \ref{hEOM} and the definitions of $S^i$, we have
\begin{equation}
	\partial_t |\vec{r}\rangle\langle\vec{r}|\sigma_1 = -\Big(\nabla\cdot\vec{j}_{\sigma_1} + \frac{\alpha}{\sqrt{g}} \big( \partial_2 (|\vec{r}\rangle\langle\vec{r}|) +  j_{1, \sigma_3} \big) \Big). \label{dtSig1} 
\end{equation} 

We integrate the expectation value of the right hand side of Eq. \ref{dtSig1} with respect to a $p_1$ eigenstate over the Gaussian pillbox B in Fig. \ref{gZcurv1var}. This causes the contribution of the first term containing the divergence operator to disappear as the longitudinal boundaries lie outside the waveguide, while the translational invariance along the longitudinal direction results in there being no difference between the contributions along the two transverse boundaries of the pillbox. Setting the integral to be 0 then gives
\begin{equation}
	\int \mathrm{d}q^2\ \langle j_{1,\sigma_3} \rangle  = -\frac{\alpha}{m} \int \mathrm{d}q^2\ \partial_2 \rho = 0. \label{j1s3} 
\end{equation}
Ithe far right hand side follows from the fact the charge density $\rho \equiv \langle \psi^\dagger\psi \rangle$ is zero on both sides of the longitudinal pillbox boundaries. 
Note that the integral on the left hand side is \textit{not}, in general, the integral of the normal spin current flowing along the longitudinal direction. The latter is  $\int \mathrm{d}q^2\ \sqrt{g}\langle j_{1,\sigma_3}\rangle $ with the extra factor of $\sqrt{g}$. This and Eq. \ref{j1s3} then together tell us that as long as the waveguide is not symmetrically curved about its transverse centre (including the case where the waveguide is flat), there will be a finite out of plane spin current flowing over longitudinal current integrated over the width of the waveguide. We show this numerically in the next section. 

Consider now the time evolution of $|\vec{r}\rangle\langle\vec{r}|\sigma_2$, we have 
\begin{eqnarray*}  
	&&\partial_t (|\vec{r}\rangle\langle\vec{r}|\sigma_2) \nonumber \\
	&=& -\nabla\cdot\vec{j}_{\sigma_2} + \frac{1}{g} \big(  (\partial_2 \ln \sqrt{g})j_{2,\sigma_2} + \frac{1}{\sqrt{g}}(\partial_2 \hat{e}_2\cdot e_3) j_{2,\sigma_3} \big)  \nonumber \\
	&&+\alpha\sqrt{g} \big( \partial_1 |\vec{r}\rangle\langle\vec{r}| -\frac{m}{g} j_{2,\sigma_3} \big)\label{dtSig2} 
\end{eqnarray*} 
where $\hat{e}_2 \equiv e_2/|e_2|$.

Taking expectation values with respect to $q^1$ eigenstates gives 
\begin{eqnarray}
0 = && \left(  -\partial_2 + (\frac{1}{g}-1) (\partial_2 \ln \sqrt{g})\right)\langle j_{2,\sigma_2} \rangle \nonumber \\
&& +\frac{1}{\sqrt{g}} \big(  (\partial_2 \hat{e}_2\cdot e_3)- m\alpha) \big) \langle j_{2,\sigma_3}  \rangle \label{dtSig2b} 
\end{eqnarray}

For completeness we state the expression for the time evolution of $|\vec{r}\rangle\langle\vec{r}|\sigma_3$. We have 
\begin{eqnarray}
	&&\partial_t (|\vec{r}\rangle\langle\vec{r}|\sigma_3) \nonumber \\
	&=&-\big(\nabla\cdot\vec{j}_{\sigma_3} + \frac{1}{g^2}(\partial_2 g)j_{2,\sigma_3} + e_2\cdot(\partial_2 \hat{e}_3)j_{2,\sigma_2} \big) - \nonumber \\
	&& m\alpha \left(\sqrt{g} j_{1,\sigma_1} + \frac{1}{\sqrt{g}} j_{2,\sigma_2}\right). \label{dtSig3}
\end{eqnarray}

\section{Numerical verification} 

We verify the predictions of our previous section by solving numerically for the eigenstates for examples of both the planar curved system, encountered in the adiabatic approximation subsection, as well as the non planar curved systems encountered in the Heisenberg EOM subsection previously. The numerical solution of the eigenstates on a curved system involves some technical subtleties due to the spatial variation of the metric tensor, which we highlight through the example of the non-planar system.

Consider a non-planar curved surface similar to the ones considered in the previous section with a finite transverse width along the $q^1$ direction, and infinite length along the $q^2$ direction.  The translational invariance along the $q^2$ direction allows the wavefunction of an eigenstate to assume the form of  $\Psi(q^1, q^2) = \exp(i k_1 q^2)\psi(q^2)$. The finite width of the segment is imposed by setting the boundary condition that the transverse wavefunction vanishes at the two edges where $\psi(q^2=0) = \psi(q^2 = W) = 0$. This boundary condition then dictates that $\psi(q^2)$ can be expanded as a sum of sine functions 
\begin{equation}
	\psi(q^2) = \sum_{n \in \mathbb{Z}^+, \sigma=(\uparrow,\downarrow) }  c_{n, \sigma} \sin\left( \frac{n\pi q^2}{W}\right)\chi_\sigma 
\end{equation}   
where $\chi_{\uparrow (\downarrow) }$ is the spin $+(-)z$ spinor. For notational simplicity we introduce the basis function $\phi_{n,\sigma} \equiv \sqrt{\frac{2}{W}}\sin\left( \frac{n \pi q^2}{W} \right)\chi_\sigma$.  

The inner product $\langle \phi_{n,\sigma}|\phi_{n', \sigma'} \rangle $ is defined as 
\begin{equation}
	\langle \phi_{n,\sigma}|\phi_{n',\sigma'} \rangle \equiv \int^W_0 \mathrm{d}q^2\ \sqrt{g}  \sin\left( \frac{n \pi q^2}{W} \right)  \sin\left( \frac{n' \pi q^2}{W} \right)\delta_{\sigma,\sigma'}. \label{innerProd} 
\end{equation}

Note that here the integration on the right hand side contains a $\sqrt{g}$ factor as the integration kernel due to the curvature of the surface, so that in general $\langle \phi_{n,\sigma}|\phi_{n',\sigma} \rangle \neq \delta_{n,n'}$. This differs from what we might be familiar with from evaluating the discrete Fourier series of a wavefucntion on a straight line. The presence of the factor of $\sqrt{g}$  in the definition of the inner product in Eq. \ref{innerProd} is necessary to ensure that $\mathbf{\tilde{H}}$ defined in the following is Hermitian and yields real values for the eigenenergy $E$.     

A numerical approximation to the Schroedinger equation may then be constructed by taking the inner product of both sides of the equality $H|\Psi\rangle = |\Psi\rangle E$ with respect to $\langle \phi_{n,\sigma}|$ -- 
\begin{eqnarray*}
	&& \langle \phi_{n, \sigma}|H|\Psi\rangle = \langle \phi_{n,\sigma}|\Psi\rangle E \\
	&\Rightarrow& \sum_{n',\sigma'}   \langle \phi_{n,\sigma}|H(k_1)|\phi_{n',\sigma'}\rangle c_{n',\sigma'} = \sum_{n',\sigma'} \langle \phi_{n,\sigma}|\phi_{n',\sigma'} \rangle c_{n',\sigma'} E \\
	&\Rightarrow& \mathbf{\tilde{H}}\mathbf{c} = \mathbf{\tilde{B}}\mathbf{c} E
\end{eqnarray*}
where in the last line $\mathbf{\tilde{H}}$ is the matrix with matrix elements $ \langle \phi_{n,\sigma}|H(k_2)|\phi_{n',\sigma'}\rangle$, $\mathbf{c}$ a column vector with elements $c_{n',\sigma'}$, and $\mathbf{\tilde{B}}$ a matrix with matrix elements $ \langle \phi_{n,\sigma}|\phi_{n',\sigma'}\rangle$.  This is then a generalized eigenvalue problem which can then be solved numerically. 

The numerical solution of the eigenstates for a planar waveguide follows largely the same ideas where similarly care must be taken to account for the position dependence of the metric tensor. 

\subsection{Planar curved waveguide} 
\begin{figure}[ht!]
\centering
\includegraphics[scale=0.5]{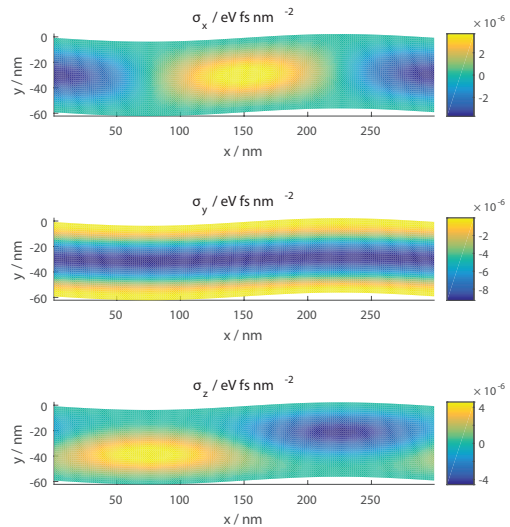}
\caption{The  charge density $\rho$, and spin densities for a finite-width sinusoidal waveguide with RSOI . }  
\label{grsCurved}
\end{figure}

To illustrate the emergence of an out of plane spin polarization on a planar curved waveguide discussed in Sect. \ref{sectAdApprox} we show in Fig. \ref{grsCurved} the charge and spin densities for one period of a periodically sinusoidal waveguide with finite width.  (The energy is chosen so that only the lowest energy transverse mode is present.) 

The state illustrated in the waveguide propagates from the left to the right of the figure.
Since this is a planar waveguide, the usual RSOI Hamiltonian on the flat $xy$ plane $\alpha (k_y\sigma_x - k_x\sigma_y)$ applies inside the waveguide. The contribution of the RSOI towards the group velocity in the $(x,y)$ directions are then $\alpha (-\sigma_y, \sigma_x)$.  ( The parameters of the waveguide and the energy can be tuned so that the contributions of the RSOI to the group velocity has the same sign as that of the kinetic energy contribution $(k_x,k_y)/m$. ) Thus, the spin $y$ accumulation is negative throughout the entire period of the waveguide because the charge carriers travelling from the left to the right of the figure always have a positive velocity in the $x$ direction. The sign of the spin $x$ accumulation alternates with the $y$ velocity of the charge carriers as they move along alternating sections of the waveguide curving towards the positive and negative $y$ directions.  In contrast, in a perfectly straight waveguide aligned along the $x$ direction $\langle k_y \rangle = 0$ for a $k_x$ eigenstate and the spin $x$ polarization is identically 0 everywhere.  The rotation of the in-plane spin polarizations in turn imply the existence of an out of plane torque which imparts an out of plane spin polarization.  Thus, the spin $z$ accumulation is positive  (negative) in those segments of the waveguide where the spins rotate clockwise (anticlockwise) as they travel from left to right.

\subsection{Non-planar arc}
We next verify the existence of an out of plane spin accumulation in a non-planar curved surface by considering a deformed c arc with a finite arc width where points on the arc are given by $\vec{r} = (r - c_\phi (\phi-\phi_0) )\hat{\phi}$ where $\phi$ is the standard cylindrical coordinate.  This represents an arc with a linear decrease in the radius with the azimuthal angle proportional to $c_\phi$. The linear decrease has been introduced to break the reflection symmetry of the arc about its middle.  In keeping with the standard convention for cylindrical coordinates we now let $z$ be the infinitely long, translationally invariant direction instead of $x$ in the previous section. The $q^1$ direction now refers to the $z$ direction, while $q^2$ remains the tangential direction along the arc.  

In order to study the effects of curvature and transverse symmetry breaking, we focus on 3 specific cases of an arc with $c_\phi = 0$, an arc with $c_\phi = 30$, and a flat waveguide.  All of these waveguides have the same transverse arc length of $60\ \mathrm{nm}$. We take $m=0.03 m_e$ and $\alpha = 0.03 \mathrm{eVnm}$ corresponding to that for an InAs heterostructure. $k_z$ is taken to be $3 \mathrm{nm}^{-1}$, 

\begin{figure}[ht!]
\centering
\includegraphics[scale=0.35]{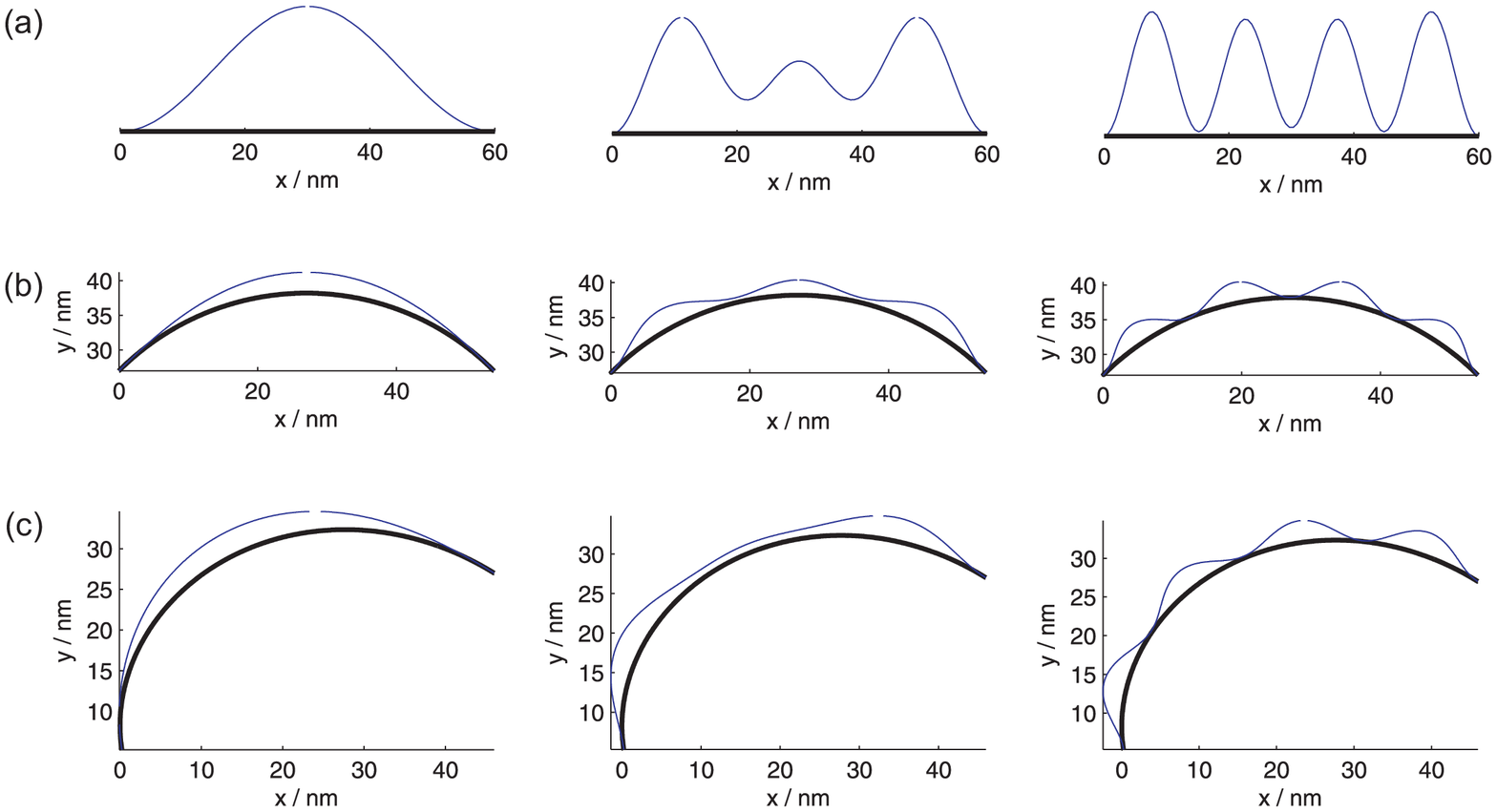}
\caption{The charge densities for the (from left to right) 1st, 5th and 7th highest energy eigenstates at the parameter ranges stated in the text for (a) a flat waveguide, (b) a perfectly symmetrical arc and (c) an asymmetric arc. The relative charge densities at each point on the waveguide are represented by the relative distance of the blue line along the normal to the point. } 
\label{gCombRho}
\end{figure}		

Despite the introduction of curvature and asymmetry, the basic shapes of the charge densities are unchanged. To numerical precision, $\langle \sigma_1(\vec{r}) \rangle$ are identically 0 everywhere for all waveguides. It was proven in Ref. \cite{PRB74_195308} that $\langle \sigma_x \rangle$ is identically 0 in a RSOI system that is translation invariant along the $q^1$ (i.e. $x$) direction and is time reversal symmetric, as is the case for our system here in the absence of a magnetization. To briefly summarize the argument there in the context of our system, the symmetry of the Hamiltonian with respect to a simultaneous transformation of $x \rightarrow -x, \sigma_y \rightarrow -\sigma_y$ implies that for a fixed value of $k_x$, we have $\langle \sigma_x(x, q^2) \rangle_{k_x} = \langle \sigma_x(-x, q^2) \rangle_{-k_x}$.  Time reversal symmetry implies that for a fixed value of $k_x$, $-\langle \sigma_x (-x, q^2) \rangle_{k_1} = \langle \sigma_x (-x, q^2)_{-k_x}$ so that $\langle \sigma_x(x, q^2) \rangle_{k_x} \rangle = -\langle \sigma_x(-x, q^2) \rangle_{k_x}$. Since the value of $\langle \sigma_x(\vec{r})\rangle$ is independent of $x$ due to translational invariance along the $x$ direction the only value it can assume is 0 everywhere. 

In agreement with the prediction of Eq. \ref{s1p2Eq} then, $\langle p^2(\vec{r}) \rangle$ is identically 0 everywhere as well. 

Consistent with $\langle \sigma_1(\vec{r}) \rangle = 0$, as we shall show shortly,  $\langle j_{2,\sigma_2}(\vec{r}) \rangle$ and  $\langle j_{2,\sigma_3}(\vec{r}) \rangle$  are identically 0. Eq. \ref{dtSig2} allows us to express $\langle j_{2,\sigma_3}\rangle$ in terms of $\langle j_{2,\sigma_2} \rangle$ and $\partial_2 \langle j_{2,\sigma_2} \rangle$. Substituting the resulting expression for $\langle j_{2,\sigma_3}\rangle$ into Eq. \ref{dtSig3} and setting $\partial_t \langle \sigma_3(\vec{r}) \rangle = 0$ gives a 2nd order differential equation in $\langle j_{2,\sigma_2}\rangle$ of the form 
\[ 
a(q^2) \partial_2^2 \langle j_{2,\sigma_2} \rangle + b(q^2) \partial_2\langle j_{2,\sigma_2} \rangle + c(q^2) \langle j_{2,\sigma_2} \rangle = \langle \sigma_1 \rangle.
\]

Note that except for the $\langle \sigma_1 \rangle$ term on the right hand side of the equal signs there are no other terms independent of $\langle j_{2,\sigma_3}\rangle$ and its $q^2$ derivatives. We have the boundary conditions that $\langle j_{2,\sigma_2} \rangle(q^2 = 0) =  \langle j_{2,\sigma_3} \rangle(q^2 = 0) = 0$ at the boundaries. Putting this into Eq. \ref{dtSig2b} gives $\partial_2 \langle j_{2,\sigma_2}\rangle(q^2 = 0) = 0$.  Therefore, if $\langle \sigma_1 \rangle = 0$ everywhere the differential equation $a(q^2) \partial_2^2 \langle j_{2,\sigma_2} \rangle + b(q^2) \partial_2\langle j_{2,\sigma_2} \rangle + c(q^2) \langle j_{2,\sigma_2} \rangle = 0$ and the boundary conditions $\partial_2 \langle j_{2,\sigma_2}\rangle(q^2 = 0) = \langle j_{2,\sigma_2}\rangle(q^2 = 0) = 0$ lead to $\langle j_{2,\sigma_3} \rangle$ being identically 0. Substituting  $\langle j_{2,\sigma_3} \rangle = \langle \sigma_1 \rangle = 0$ into Eq. \ref{dtSig2} in turn leads to $\langle j_{2,\sigma_2} \rangle$ being identically equal to 0 as well.  The expectation values of all the spin currents that appear in Eq. \ref{dtSig2} are hence identically 0 everywhere.

\begin{figure}[ht!]
\centering
\includegraphics[scale=0.3]{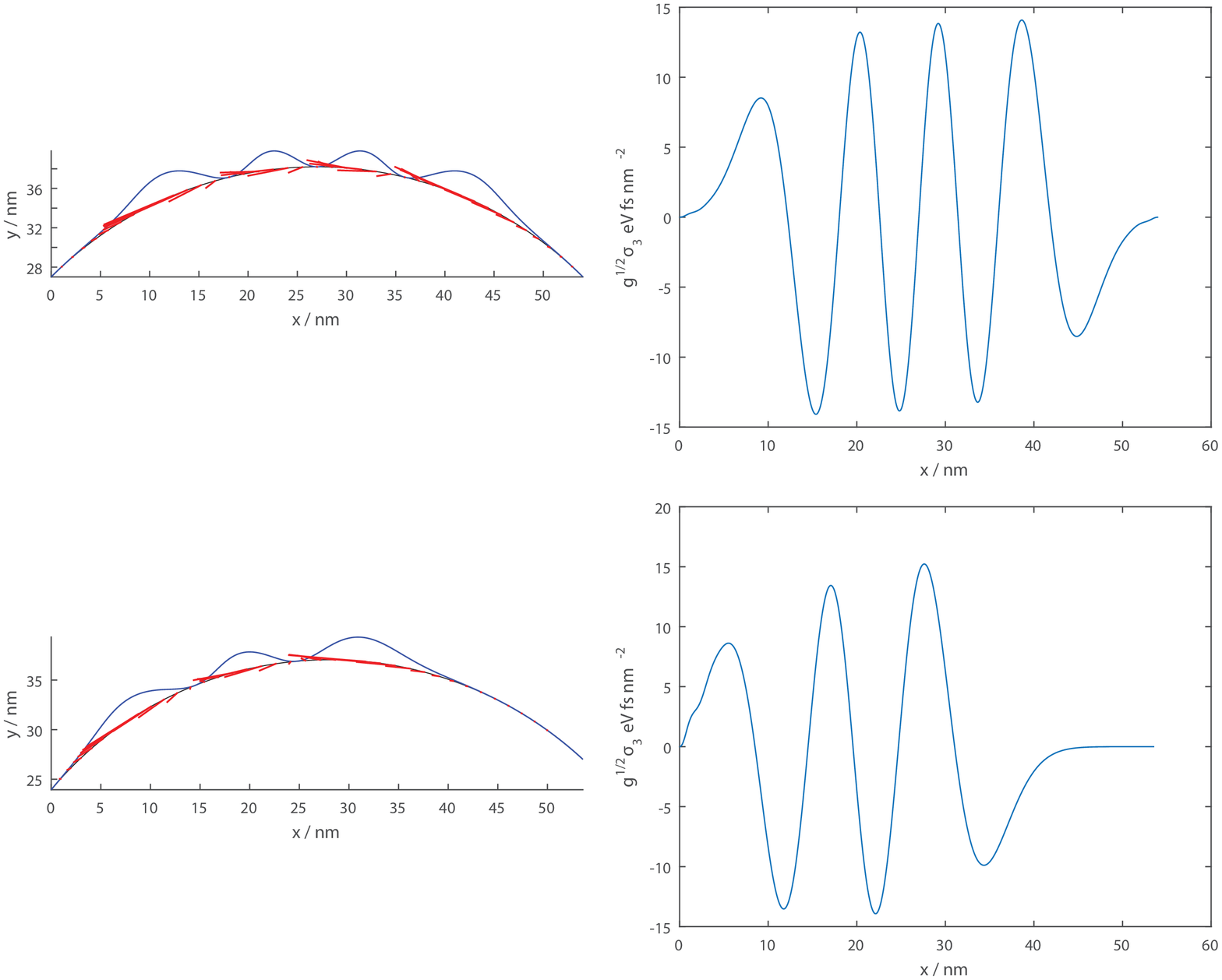}
\caption{The left panels show the directions and relative magnitudes of the spin densities at various points on (a) the symmetrical arc and (b) the asymmetrical arc. The right panels show plots of $\sqrt{g}\langle\sigma_3(\vec{r})\rangle$.   } 
\label{gCombSig3}
\end{figure}		

However, as predicted in the discussion after Eq. \ref{dtSig1}, Fig. \ref{gCombSig3} shows that the presence of asymmetry breaks the antisymmetry of the out of plane spin current flowing along the longitudinal direction. There is thus a net longitudinal out of plane spin current after integrating over the entire width of the waveguide. 

\section{Conclusion}
In this work we derived the RSOI Hamiltonian on arbitrarily curved surfaces. With this Hamiltonian, we used the adiabatic approximation to show that analogous to the acceleration of the charge carriers by an electric field, the change in momentum direction of the charge carriers as they move along a gently curved planar waveguide lying flat on the $xy$ plane leads to an out of plane spin accumulation. We then used the Heisenberg equations of motion to analyze the relationships between spin currents and accumulations on curved waveguides. We showed that for the eigenstates of a time independent RSOI Hamiltonian, the spin torque on the charge carriers due to the SOI field can be understood as being balanced off by the divergence in the spin currents and additional terms due to curvature on a curved waveguide. We also showed that the equations of motion predict the existence of out of plane spin polarization on asymmetrically curved, non-planar waveguides. We then described the numerical solution of the eigenstates on curved waveguides, and presented numerical results to verify our earlier analytic predictions.

\section{Acknowledgements}
We thank the MOE Tier II grant MOE2013-T2-2-125 (NUS Grant No. R-263-000-B10-112), and the National Research Foundation of Singapore under the CRP Program ''Next Generation Spin Torque Memories: From Fundamental Physics to Applications'' NRF-CRP9-2013-01 for financial support.

\section{Appendix} 

We also need an expression for $[ |\vec{r}\rangle O \langle\vec{r}|, \frac{p^2}{2m} ]$ where $O$ is an operator which commutes with $|\vec{r}\rangle\langle\vec{r}|$.  

We consider
\begin{eqnarray*}
	&&\langle \Psi| [ |\vec{r}\rangle O \langle\vec{r}|, p^2 ] |\Phi\rangle\\
	&=& -\frac{1}{\sqrt{g}} ( \Psi^* O\partial_i(\sqrt{g} g^{ij} \partial_j\Phi) - \partial_i (\sqrt{g}g^{ij}\partial_j\Psi^*) O\Phi)
\end{eqnarray*}

Now
\begin{eqnarray*}
&& \Psi^* O\partial_i(\sqrt{g} g^{ij} \partial_j \Phi) - (\partial_i (\sqrt{g}g^{ij} \partial_j \Psi^*) O\Phi\\
&=&\sqrt{g}\Big( ( \Psi^* O)\big( \partial_i(g^{ij}\partial_j\Phi) + g^{ij}(\partial_i \ln \sqrt{g})(\partial_j\Phi)\big)  \\
&&- \big(\partial_i(g^{ij}\partial_j\Psi^*) + g^{ij}(\partial_i\ln \sqrt{g})(\partial_j\Psi^*)\big)(O\Phi))\Big)\\
&=&\sqrt{g}\Big( (\partial_i + (\partial_i \ln\sqrt{g}))(\Psi^*Og^{ij}\partial_j\Phi - (\partial_j\Psi^*)Og^{ij}\Phi) \\
&& - \partial_i (\Psi^*O)g^{ij}\partial_j\Phi + g^{ij}(\partial_j \Psi^*)\partial_i(O\Phi)\Big).
\end{eqnarray*}

In the first line of the last equality sign above, we have, 
\begin{eqnarray*}
	&&g^{ij} (\Psi^*O\partial_j\Phi - (\partial_j\Psi^*)O\Phi)\\
	&=& i g^{ij}(\Psi^*O (-i\partial_j\Phi) + (i\partial_j\Psi^*)O\Phi)\\
	&=& \frac{i}{\hbar} g^{ij}\langle \Psi| \{ |\vec{r}\rangle O \langle\vec{r}|,p_j \} |\Phi\rangle 
\end{eqnarray*}

while on the second line we have 
\begin{eqnarray*}
&& - \partial_i (\Psi^*O)g^{ij}\partial_j\Phi + g^{ij}(\partial_j \Psi^*)\partial_i(O\Phi) \\
&=& g^{ij}(\Psi^*(-\partial_i O)\partial_j\Phi + \partial_j \Psi^*(\partial_i O)\Phi) \\
&=& (-i) g^{ij}(\Psi^*(\partial_i O)(-i\partial_j\Phi) + (i\partial_j\Psi^*)\partial_i O\Phi) \\
&=& -\frac{i}{\hbar} g^{ij} \langle \Psi| \{|\vec{r}\rangle\partial_i O\langle\vec{r}|,p_j\} |\Phi\rangle
\end{eqnarray*}

so that putting everything back together, we have 
\begin{eqnarray}
	&&-i[|\vec{r}\rangle O \langle \vec{r}|, p^2] =  \nonumber \\
	&& -\Big( \big((\partial_i + (\partial_i \ln\sqrt{g}))\{|\vec{r}\rangle O \langle\vec{r}|, p^i \}\big) \nonumber \\
	&\ & - \{|\vec{r}\rangle\partial_i O\langle\vec{r}|,p^i\} \Big).
\label{AKEeom}
\end{eqnarray}

\end{document}